\documentclass[letterpaper, 10 pt, conference]{ieeeconf}  

\IEEEoverridecommandlockouts                              
\usepackage{graphicx}       
\usepackage{amssymb}
\usepackage{amsmath}
\usepackage{multirow}
\usepackage{hyperref}
\hypersetup{
    colorlinks=true,
    linkcolor=blue,
    filecolor=magenta,      
    urlcolor=cyan,
}

\title{\LARGE \bf
Dynamic-Aware Spatio-temporal Representation Learning for Dynamic MRI Reconstruction
}

\author{Dayoung Baik, and Jaejun Yoo}

\begin{document}

\maketitle
\thispagestyle{empty}
\pagestyle{empty}

\begin{abstract}

Dynamic MRI reconstruction, one of inverse problems, has seen a surge by the use of deep learning techniques. Especially, the practical difficulty of obtaining ground truth data has led to the emergence of unsupervised learning approaches. A recent promising method among them is implicit neural representation (INR), which defines the data as a continuous function that maps coordinate values to the corresponding signal values. This allows for filling in missing information only with incomplete measurements and solving the inverse problem effectively. Nevertheless, previous works incorporating this method have faced drawbacks such as long optimization time and the need for extensive hyperparameter tuning. To address these issues, we propose Dynamic-Aware INR (DA-INR), an INR-based model for dynamic MRI reconstruction that captures the spatial and temporal continuity of dynamic MRI data in the image domain and explicitly incorporates the temporal redundancy of the data into the model structure. As a result, DA-INR outperforms other models in reconstruction quality even at extreme undersampling ratios while significantly reducing optimization time and requiring minimal hyperparameter tuning. Our code is available at \href{https://github.com/9B8DY6/DA_INR}{here}.

\end{abstract}

\section{INTRODUCTION}

Dynamic Magnetic Resonance Imaging (MRI) captures sequential images of moving organs, such as the heart, while Dynamic Contrast Enhanced (DCE) MRI monitors temporal changes in in-vivo drug effects on vasculature. Due to the slow acquisition speed of MRI, only partial data can be collected per frame, leading to a trade-off between spatial and temporal resolution. 
Recent approaches have accelerated data acquisition while maintaining image quality by exploiting sparsity in the spatial and temporal domains~\cite{ktfocuss, lowrankreg_, grasp}. Early deep learning methods~\cite{supervised, supervised_2, supervised_3, supervised_4} applied supervised learning, but required large amounts of paired undersampled and fully sampled data, limiting their practicality.

To address this, unsupervised learning methods have emerged, leveraging inherent priors in Convolutional Neural Networks (CNNs)~\cite{yoo2021timedependent} and Implicit Neural Representations (INRs)~\cite{huang2023neural, kunz2024implicit}. CNN-based approaches, such as~\cite{yoo2021timedependent}, exploit the structural prior of randomly initialized CNNs to capture low-level image statistics, which serves as an implicit regularization across all frames. However, the use of discrete grid representations in CNN constrains their ability to fully capture the continuous nature of dynamic MRI data. In contrast, INR-based methods~\cite{huang2023neural, kunz2024implicit} represent dynamic MRI data as a continuous neural function in both spatial and temporal dimensions. By optimizing this function using spatio-temporal coordinates as inputs and predicting the corresponding values based on the available measurements, INR models effectively infer missing information during the reconstruction process, enabling full data recovery. 

Specifically, Neural Implicit \textit{k}-space (NIK)~\cite{huang2023neural} introduced learning neural representation in the frequency domain to avoid regridding loss, but aliasing artifacts remained evident in the reconstructed images. The Fourier feature MLP (FMLP)~\cite{kunz2024implicit} used a Fourier feature encoder~\cite{fourierfeatures} without requiring explicit regularization terms and showed superior performance over previous methods. However, a common drawback across all these approaches is the lengthy optimization process, which can take several hours to an entire day for networks to converge. More recent work~\cite{feng2023spatiotemporal} replaces the Fourier feature encoder with a hash encoder~\cite{mueller2022instant} to achieve faster convergence. However, this approach remains time-intensive due to complexity of tuning hyperparameters for the hash encoders and regularization terms required for spatial and temporal consistency. Moreover, the results are highly sensitive to the weighting of these regularization terms.


To address these challenges, we propose Dynamic-Aware INR (DA-INR), which explicitly models the temporal redundancy inherent in dynamic MRI data, inspired by D-NeRF~\cite{d-nerf}. It circumvents the need for manual weighting of regularization terms by making canonical space play as a regularization role to the other frames during optimization. Thus, it enables more stable convergence than a hash encoder alone. As a result, DA-INR not only simplifies the training process, but also enhances adaptability to diverse undersampling conditions and data complexities, offering an efficient solution for dynamic MRI reconstruction. 
\begin{itemize}
\item We propose Dynamic-Aware Implicit Neural Representation (DA-INR) that is explicitly designed to model temporal redundancy in dynamic MRI data. Accordingly, Our model enables stable convergence even without any regularization term.
\item Our model does not depend on regularization terms and is not sensitive to hyperparameter value variations, so we do not need to fine-tune hyperparameters for regularization terms or etc.
\item We show that DA-INR achieves state-of-the-art performance on cardiac cine and DCE liver datasets under various conditions on dynamic MRI reconstruction. It accelerates optimization by $4.63\times-9.78\times$ and reduces GPU memory usage by $2.15\times-2.89\times$ compared to the existing methods.
\end{itemize}

\section{Background}

\subsection{Dynamic MRI reconstruction}
In dynamic MRI, the relationship between an undersampled $(k,t)$-space data and a discrete reconstructed image sequence matrix can be represented as below:
\begin{equation}\label{eq:dynamicmri_fft}
\begin{split}
    m_c &=F_uS_cd.
\end{split}
\end{equation}
Here, $S_c \in \mathbb{C}^{(N\times N) \times (N\times N)}$ is a diagonal matrix representing $c^{th}$ coil sensitivity map $(1 \le c \le C)$, $m_c \in \mathbb{C}^{N\times M \times \tau}$ is a $c^{th}$ coil undersampled $(k,t)$ space data, $d \in \mathbb{C}^{N\times N \times \tau}$ is a discrete reconstructed image sequence matrix, $\tau$ is temporal length of image sequence $d$, $N$ is an image size, $M$ is the number of undersampled lines per frame $(N>M)$, and $C$ is the total number of coil channels. $F_u \in \mathbb{C}^{(N\times M) \times (N\times N)}$ denotes an undersampled Fourier operator that simulates undersampled data acquisition. If the \textit{k}-space data is sampled in golden angle radial-way, $F_u$ denotes Non-uniform Fast Fourier Transform (NuFFT) operator. 

Reconstructing image $d$ from the undersampled $(k,t)$-space data $m_c$ is actually one of the ill-posed (inverse) problems, and the optimization process is formulated as:
\begin{equation}\label{eq:dynamicmri_recon_optimization}
\begin{split}
    \underset{d}{argmin} \sum^C_{c=1}\underbrace{||F_uS_cd-m_c||^2_2}_\text{\clap{Data consistency}} + \lambda\mathbf{R}(d),
\end{split}
\end{equation}
where $\mathbf{R}(d)$ is an explicit regularizer which is based on prior for more stable optimal reach and $\lambda$ is its weight, a hyperparameter. $\mathbf{R}(d)$ can be low rank regularization~\cite{lowrankreg_}, temporal total variation (TV) regularization~\cite{grasp}, or both etc. 

\subsection{Implicit Neural Representation (INR)}
Implicit Neural Representation (INR) is a novel scheme in deep learning of which Multi-Perceptron Layers (MLP) is a parameterization of signal or data. In contrast to CNN which has discrete nature according to discontinuity of data, e.g. discrete image or voxel, INR represents data as continuous neural function with continuous coordinates taken as inputs. When the coordinates are given to INR, it outputs the corresponding values, such as RGB or density. Given data $\mathbf{d} \in \mathbb{C}^{H\times W}$, an optimized neural function $f_\theta$ represents it as $\mathbf{d}_\theta$ which is formulated as:
\begin{equation}
\begin{split}
    \mathbf{d}_\theta &= \begin{bmatrix}
f_\theta (1,1) & \cdots & f_\theta (1,W)\\
\vdots & \cdots & \vdots \\
f_\theta (H,1) & \cdots & f_\theta (H,W)\\
\end{bmatrix}
\end{split}
\end{equation}

Recently, INR has emerged as a new solver for dynamic MRI reconstruction by benefiting from the data continuity prior imposed in INR~\cite{huang2023neural, kunz2024implicit, feng2023spatiotemporal}. The parameters $\theta$ of an untrained neural network $f_\theta$ are optimized to produce an image sequence $d_\theta$ that is consistent with the $c^{th}$ coil measurement $m_c$. The problem being solved is formalized as:
\begin{equation}\label{eq:lossterm_general}
\begin{split}
    \underset{\theta}{argmin} \sum^C_{c=1}||F_uS_cd_\theta - m_c||^2_2 + \lambda\mathbf{R}(d_\theta).
\end{split}
\end{equation}

\begin{figure*}[!t]
\centerline{\includegraphics[width=\textwidth]{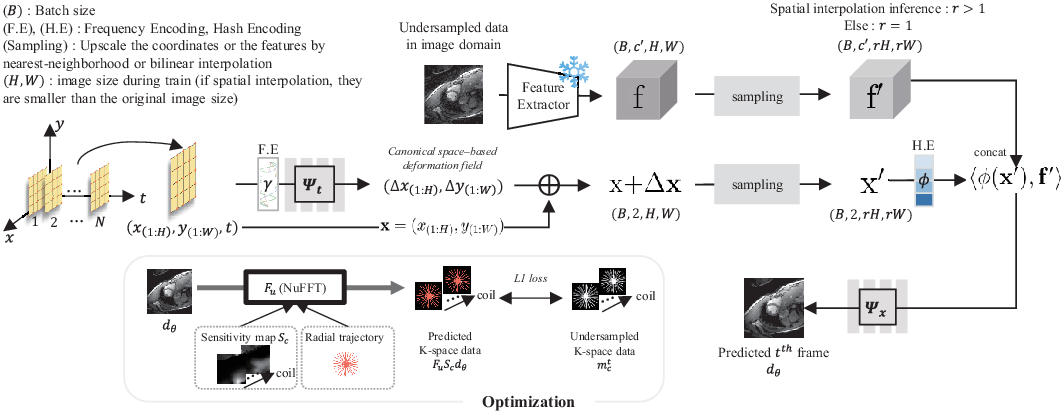}}
\caption{DA-INR model architecture. A deformation network $\Psi_t$ takes a spatio-temporal coordinate $(x,y,t)$ as input to output deformation field $\Delta \mathbf{x}=(\Delta x, \Delta y)$ based on a canonical space. A pretrained feature extractor extracts features from an undersampled data in the image domain. A canonical network $\Psi_x$ takes the deformed coordinate $\mathbf{x}'$ and the features $\mathbf{f}'$ to predict $t^{th}$ frame in the image domain, $d_\theta$. These two models are optimized by L1 loss computation in the frequency domain with Non-uniform Fast Fourier Transform (NuFFT).} \label{fig:Ours_method}
\end{figure*}

\section{Method}
\subsection{Dynamic-aware HashINR}
We propose Dynamic-aware HashINR (DA-INR), which accelerates the optimization process in dynamic MRI reconstruction through a novel dynamic hash encoding scheme. 
In this section, we provide an overview of (1) the overall workflow, (2) the core components of the framework, (3) the hash encoding method, and (4) the optimization strategy. 

\subsubsection{Overall workflow}
DA-INR consists of three learning stages (Fig.\ref{fig:Ours_method}). The framework operates within the canonical space, which serves as a reference coordinate system that captures the static structure of the dynamic MRI data. The input coordinate $(x,y,t)$ is encoded by frequency encoding~\cite{nerf} and passed into the deformation network $\Psi_t$ which outputs the deformation field $(\Delta x, \Delta y)$ based on the canonical space. The pretrained feature extractor extracts image features from the undersampled data in the image domain. Then, the canonical network $\Psi_x$ takes the deformed coordinate $(x+\Delta x, y+\Delta y)$ and the image features as input and predicts the corresponding value within the canonical space. 

\subsubsection{Deformation network}
The deformation network $\Psi_t$ estimates the deformation field between cells at a specific time $t$ and cells in the canonical space. More specifically, given the input coordinate $\mathbf{x} = (x,y)$ at time $t$, $\Psi_t$ predicts the deformation field $\Delta \mathbf{x}$ to transform the cell position $(x,y)$ to the cell position $(x+\Delta x, y+\Delta y)$ in the canonical space. Before going into $\Psi_t$, $\mathbf{x}$ and $t$ is encoded by the frequency encoding, $\gamma (p)=\left<(sin(2^i\pi p),cos(2^i\pi p))\right>^I_0$~\cite{nerf}. It is applied to each component of the input coordinate with $I=10$ and the time component with $I=6$. $\Psi_t$ is defined as:
\begin{equation}\label{eq:deformationnet_ours}
\begin{split}
    \Psi_t(\gamma(\mathbf{x})_{I=10}),\gamma(t)_{I=6}) &=
    \begin{cases}
    \Delta \mathbf{x}, & \mbox{if } t \not= 0 \\
    0, & \mbox{if } t = 0
    \end{cases}
\end{split}
\end{equation}

\subsubsection{Feature extraction}
We use the pretrained image feature extractor~\cite{mdsr} to obtain additional information for dynamic MRI reconstruction. 
The frozen feature extractor takes the undersampled data in the image domain at time $t$ (spatial interpolation, reconstruction) or of two neighboring frames of time $t$ (temporal interpolation) as input and outputs the image features $\mathbf{f} \in \mathbb{R}^{c' \times H \times W}$ whose size is the same as the input frame. $c'$ is the size of the channel dimension. We upscale $\mathbf{f}$ to $\mathbf{f}' \in \mathbb{R}^{c' \times rH \times rW}$ by bilinear interpolation, and $\mathbf{x} + \Delta \mathbf{x}$ to $\mathbf{x}' \in \mathbb{R}^{2 \times rH \times rW}$ by nearest-neighborhood interpolation based on the scale ratio $r$. During optimization, $r$ is fixed as 1, $r=1$. During inference of spatial interpolation, $r$ is bigger than 1, $r>1$.

\subsubsection{Canonical network}
The canonical network $\Psi_x$ predicts the corresponding image intensity value in the canonical space, given the resampled deformed coordinate $\mathbf{x}'$ and the image features $\mathbf{f}'$. The input $\mathbf{x}'$ is first encoded by a hash encoder $\phi$~\cite{mueller2022instant} and is concatenated with $\mathbf{f}'$ in the channel dimension. Then, they are fed into $\Psi_x$ to output the corresponding image intensity value in the cell position of the canonical space. The canonical network $\Psi_x$ is defined as:
\begin{equation}\label{eq:canonicalnet}
\begin{split}
    \Psi_x(\langle \phi(\mathbf{x}') ,\mathbf{f}' \rangle) &= (Re, Im),
\end{split}
\end{equation}
where $Re$ and $Im$ denote real and imaginary components of the cell position $\mathbf{x}'$ in the complex-valued image, respectively, and $\langle \cdot,\cdot \rangle$ means channel-wise concatenation. The final output image $d_\theta \in \mathbb{C}^{rH\times rW}$ at time $t$ is defined as:
\begin{equation}\label{eq:outputimage}
\begin{split}
    d_\theta &= \begin{bmatrix}
f_\theta (1,1,t) & \cdots & f_\theta (1,rW,t)\\
\vdots & \cdots & \vdots \\
f_\theta (rH,1,t) & \cdots & f_\theta (rH,rW,t)\\
\end{bmatrix},
\end{split}
\end{equation}
where $f_\theta$ is DA-INR, and $rH$ and $rW$ are height and width of the final output image.

\subsubsection{Hash encoding}
The hash encoder $\phi$ exploits a multi-resolution grid to extract variable features from a viewpoint of various resolutions, and a hash table to store trainable feature vectors. The hash table is arranged into $L$ levels, each containing up to $T$ feature vectors with dimension $F$. Each level corresponds to respective resolution of the grid and each feature vector in the hash table corresponds to that in the edge of the grid. It is assigned by spatial hashing function~\cite{Teschner2003OptimizedSH}. 
Depending on where the input coordinate $\mathbf{x}'$ is located in the grids, the corresponding feature vector $\phi_l \in R^F, l=0,1,...,L-1$ at $\mathbf{x}'$ is linearly interpolated between the feature vectors at adjacent edges in all levels. All vectors, interpolated for all levels, are concatenated as $\phi = concat(\phi_1,\phi_2,...,\phi_L)$ and fed into neural network to estimate the corresponding value. 
The resolution of the grid, $N_l$, follows a geometric series, $N_l = \lfloor N_{min} \cdot b^l \rfloor$, where $N_{min}$ is the coarsest resolution and $b$ is growth ratio. As level $l$ increases, it focuses on the finer details of the data. $N_{max}$ is the finest resolution and there is a geometric progression between the coarsest and finest resolution with $b$.
\begin{equation}\label{eq:growthratio}
\begin{split}
b := \exp(\frac{\ln(N_{max}) - \ln(N_{min})}{L-1}).
\end{split}
\end{equation}
$L,T,F,N_{min}$, and $b$ are trade-off hyper-parameters between memory, performance and quality.

\subsubsection{Optimization}
We only use L1 loss as data-consistency for optimization. The final loss $\mathcal{L}$ is defined as follows:
\begin{equation}\label{eq:lossterm_ours}
\begin{split}
    \mathcal{L} &= \sum^C_{c=1}||F_uS_cd_\theta - m_c^t||^1_1,
\end{split}
\end{equation}
where $d_\theta$ is the reconstructed image by DA-INR defined in Eq. \ref{eq:outputimage}, $m_c^t$ is a $c^{th}$ coil golden-angle radial undersampled \textit{k}-space data at time $t$. $F_u$ is NuFFT operator with a given radial trajectory and multi-coil sensitivity map. 

\begin{figure}[tb!]
\centering
\includegraphics[width=\linewidth]{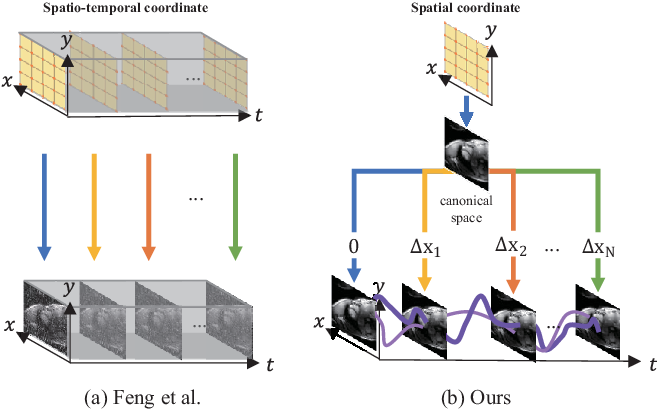}
\caption{Visual comparison of (a) Feng et al. and (b) DA-INR. Feng et al. represents dynamic MRI data as a three dimensional measurement with $x$-axis, $y$-axis, and $t$-axis and each coordinate $(x,y,t)$ directly maps to each cell in the image at time $t$. In contrast, in DA-INR, the cells of the image in the canonical space plays a regularization role to those of all other frames. The purplish lines between frame-by-frame in (b) indicate that DA-INR is continuous in time, but does not merely represent dynamic MRI data as 3D mass.} \label{fig:fitting_comp.png}
\end{figure}
\subsection{Difference between Feng et al. and DA-INR for encoding temporal redundancy} \label{subsec:effectsoftemporalredundancy}
Feng et al.~\cite{feng2023spatiotemporal} learns to map $(x,y,t)$ directly to the corresponding value in the image domain based on the following optimization on undersampled \textit{k}-space data (Fig.\ref{fig:fitting_comp.png} (a)):
\begin{equation}\label{eq:hashinr_opt}
\begin{split}
    \underset{\theta}{argmin} \sum^C_{c=1} ||F_uS_cd_\theta-m_c||^2_2 + \lambda_S ||TV_t(d_\theta)||_1 + \lambda_L||d_\theta||_*.
\end{split}
\end{equation}
This equation is based on Eq.~\ref{eq:lossterm_general}. $\mathbf{R}(d_\theta)$ is composed of temporal TV and low rank regularization terms. $TV_t(\cdot)$ denotes a temporal TV operator and $||\cdot||_*$ denotes a sequent operator of singular value decomposition and sum of singular values. $\lambda_S$ and $\lambda_L$ are the weights for the temporal TV and the low rank regularization, respectively. In contrast, DA-INR fits all the frames to the shared canonical space, constraining the solution space through the canonical network $\Psi_x$. The fitting process in DA-INR can be interpreted as mapping $(x,y)$ in the canonical space to the corresponding intensity values, while other frames retrieve these signal values at their cell points based on the deformation field $[\Delta x_k]_{k=1}^N$, which accounts for the temporal difference (Fig.\ref{fig:fitting_comp.png} (b)). Since the canonical space is shared across all frames, this design induces interconnection within the image sequence, allowing the model to focus on signals in the canonical space and enabling faster compensation for artifacts caused by undersampling in the other frames. 

\section{Experiments}
\begin{figure}[tb!]
\centering
\includegraphics[width=\linewidth]{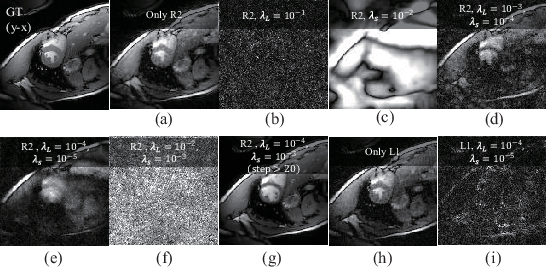}
\caption{Sensitivity analysis of Feng et al.~\cite{feng2023spatiotemporal} on cardiac cine data. ``R2'' denotes the use of relative L2 loss~\cite{relativel2loss} for data-consistency optimization, while ``L1'' refers to the use of L1 loss. $\lambda_L$ and $\lambda_S$ are weights for low rank and temporal TV regularization, respectively. Lastly, ``$step>20$'' denotes that temporal TV regularization is turned on after training step $20$.}
\label{fig:hyperparam_tuning_initialexp}
\end{figure}

\subsection{Baseline methods}
We compare our method against Non-uniform Fast Fourier Transform (NuFFT), GRASP~\cite{grasp}, TD-DIP~\cite{yoo2021timedependent}, and the method proposed by Feng et al.~\cite{feng2023spatiotemporal}. 
NuFFT represents the results obtained by directly transformed zero-filling data in the \textit{k}-space domain\footnote{\label{nufft1}https://github.com/dfm/python-nufft}.
GRASP~\cite{grasp} is a compressed sensing (CS) method that applies explicit temporal TV regularization based on the sparsity prior of the image in the temporal TV domain. TD-DIP~\cite{yoo2021timedependent} is an unsupervised method that leverages the implicit structural regularization inherent in CNNs to learn a mapping from a low-dimensional time-dependent vector at a specific time step to the corresponding high-dimensional frame. 
Feng et al.~\cite{feng2023spatiotemporal} is an INR-based method with hash encoding that incorporates explicit temporal TV and low-rank regularization terms for optimization. 
They optimize their model with Eq.~\ref{eq:hashinr_opt} and set different values on the weights of the regularization terms according to data types and acceleration factors. However, due to the sensitivity of the results to minor variations in regularization weights and their dependence on the data types (Fig.~\ref{fig:hyperparam_tuning_initialexp}), we decide to reproduce this method \textit{without} these regularization terms. This adaptation is referred to as \textit{HashINR} in our paper. 

\subsection{Datasets}
\subsubsection{Retrospective cardiac cine data}
Cardiac cine data was obtained using a 3T whole-body MRI scanner (Siemens Tim Trio) equipped with a 32-element cardiac coil array. The acquisition utilized a bSSFP sequence with prospective cardiac gating. The imaging parameters were as follows: FOV $=(300\times300)$ mm\textsuperscript{2}, acquisition matrix $=(128 \times 128)$, TE/TR=$1.37/2.7$ ms, receiver bandwidth = 1184 Hz/pixel, and flip angle $= 40^{\circ}$. The number of frames was 23 and the temporal resolution was 43.2 ms. 
The full-sampled \textit{k}-space data is used as ground truth (GT). To simulate a retrospective undersampling pattern, we adopt a 2D golden-angle radial acquisition scheme, where the spokes repeatedly traverse the center of \textit{k}-space, rotating with a step of $111.25^{\circ}$. It is applied to ground truth with multi-coil NuFFT to obtain the undersampled radial trajectories of Fibonacci numbers~\cite{4242}. 

\subsubsection{Dynamic Contrast-Enhanced (DCE) liver data}
The DCE liver MRI scan was conducted on a healthy volunteer using axial orientation and breath-holding techniques by a whole-body 3 Tesla MRI system (MAGNETOM Verio/Avanto, Siemens AG, Erlangen, Germany), employing a combination of body-matrix and 12-element spine coil array. For data acquisition, a radial stack-of-stars 3D Fast Low Angle Shot (FLASH) pulse sequence with golden-angle ordering was utilized. The imaging parameters were as follows: FOV = $370 \times 370$ mm\textsuperscript{2}, TR/TE=3.83 ms/1.71 ms, acquisition matrix = $384 \times 384$, total spoke for one slice $=600$. 
In our experiments, 34 consecutive spokes are used for each frame, which provides 17 temporal frames. The reconstructed image matrix for each frame is 384$ \times $384. The acceleration factor $(AF)$ is $384/34=11.3$.

\subsection{Performance evaluation}
For cardiac cine data, we use Peak Signal-to-Noise Ratio (PSNR) and structural similarity index (SSIM) as evaluation metrics, both calculated frame-by-frame. The ground truth images and the estimated outputs are normalized to the range $[0,1]$ based on the maximum and minimum values of each sequence. For DCE liver data, we conduct a visual comparison and assess temporal fidelity based on the signal intensity in the region of interest (ROI), as no ground truth images are available. The ROIs for the aorta (AO) and portal vein (PV) are manually drawn for each signal intensity flowmap. We use NuFFT as a reference because the contrast changes can be preserved due to the average signal intensity over a large ROI. 
We test the performance of each method with 21, 13, and 5 spokes per frame $(AF = 6.1, 9.8, 25.6)$ on cardiac cine data, and with 34 spokes per frame $(AF = 11.3)$ on DCE liver data. 

\subsection{Implementation details}
For the deformation network and the canonical network, we use a small MLP with five hidden layers, each comprising 64 neurons, followed by a ReLU activation function. The MLP of the canonical network predicts two channels representing the real and imaginary components of the complex-valued MRI images, with no activation function applied to the final layer. During the optimization process, all spatio-temporal coordinates are assembled in a single batch and normalized to the range $[-1, 1]$. The batch size is set to 1. We use the AdamW~\cite{AdamW} optimizer with a learning rate of 0.001, $\beta_1=0.9$, $\beta_2=0.999$, and $\epsilon=10^{-8}$ during optimization. The non-Cartesian Fourier undersampling operation is executed using the NuFFT package\footref{nufft1}, facilitating rapid computation and gradient backpropagation on a GPU. Feng et al.~\cite{feng2023spatiotemporal} tunes the hyperparameters $L,T,F,N_{min}$, and $b$ of its hash encoder according to each data type and AF, while we use consistent values across all data types and AFs as $L=16,T=2^{19},F=2,N_{min}=16$, and $b=2$.

\begin{figure*}[htb!]
\centering
\includegraphics[width=\textwidth]{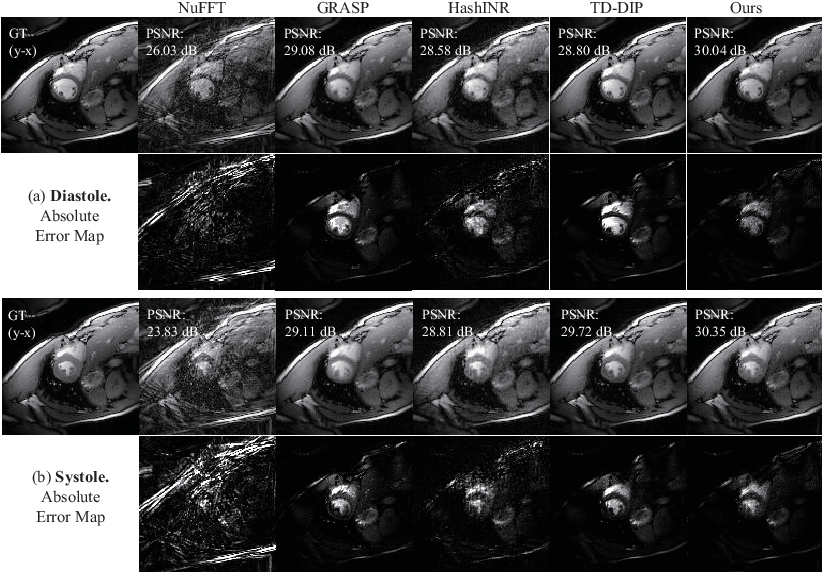}
\caption{Visual comparisons between results of $AF=9.8$ in cardiac cine data reconstruction at diastole and systole. The upper row is the reconstruction output in the $(y-x)$ domain for each method and the below row is the absolute error map between ground truth and the reconstructed output of each method. PSNR values are specific to each frame.
}
\label{fig:errormapvis_13}
\end{figure*}

\begin{figure*}[htb!]
\centering
\includegraphics[width=\textwidth]{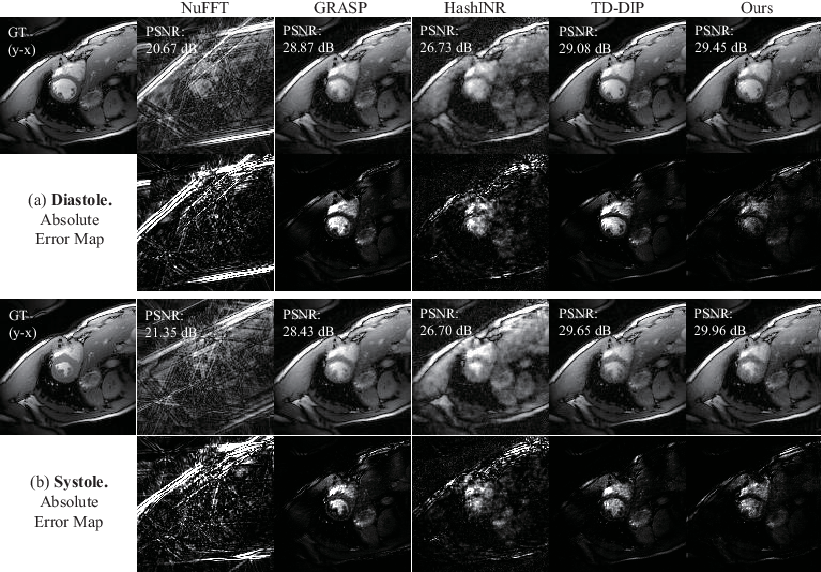}%
\caption{
Visual comparisons between results of $AF=25.6$ in cardiac cine data reconstruction at diastole and systole. The upper row is the reconstruction output in the $(y-x)$ domain for each method and the below row is the absolute error map between ground truth and the reconstructed output of each method. PSNR values are specific to each frame.
}
\label{fig:errormapvis_5}
\end{figure*}

\section{Results}
\subsection{Retrospective cardiac cine data}
\begin{figure*}[htb!]
\centering
\includegraphics[width=\textwidth]{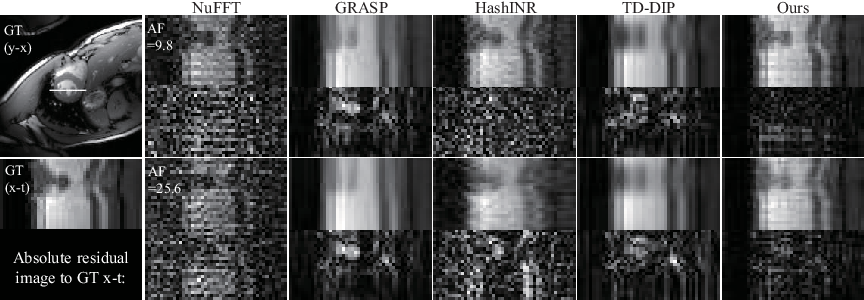}
\caption{Qualitative results of cardiac cine data reconstruction in the $(x-t)$ domain. A white line at the heart region indicates the cross section that is visualized.}
\label{fig:cardiac_ytdomain}
\end{figure*}

Fig.~\ref{fig:errormapvis_13} and Fig.~\ref{fig:errormapvis_5} present the visual comparisons between our method and the existing methods for cardiac cine data reconstruction at $AF=9.8$ and $AF=25.6$, respectively. We evaluate the reconstructed frames for each method during the diastolic and systolic phases. When undersampling is performed with only five spokes per frame ($AF = 25.6$), NuFFT struggles to accurately capture the cardiac structure. For HashINR, reconstruction quality improves as the number of sampled spokes increases; however, residual noise persists even at $AF = 9.8$. While GRASP sometimes achieves high PSNR, it struggles to reconstruct fine structural details, such as the shape of the papillary muscle, under both undersampling ratios. TD-DIP shows limitations in accurately capturing the precise contraction of the papillary muscle during the systolic phase at $AF=25.6$. In contrast, our method closely approximates the ground truth, achieving high fidelity in both the diastolic and systolic phases. The quantitative error---computed as the sum of the squared differences between the reconstructed and ground truth images---is the smallest among all methods. Fig.~\ref{fig:cardiac_ytdomain} presents the qualitative reconstruction results for cardiac cine data in the $(x-t)$ domain. While TD-DIP and GRASP produce satisfactory results in the $(y-x)$ domain, their performance deteriorates in the $(x-t)$ domain. HashINR achieves reconstructions in the $(x-t)$ domain comparable to the ground truth at $AF = 9.8$, though with some noise, and suffers from pronounced blurring along the time dimension at $AF = 25.6$. In contrast, our method demonstrates superior reconstruction quality, with reduced noise and minimal blurring.

\begin{table}
\centering
\caption{Quantitative results of cardiac cine data reconstruction. We compare ours to NuFFT, GRASP, TD-DIP, and HashINR at $AF=25.6$ and $AF=9.8$.}
\label{tab:cardiac_5_13}
\resizebox{\linewidth}{!}{
\begin{tabular}{cccc}
\hline
\multicolumn{1}{l}{Method} & Undersampling ratio        & PSNR (dB)      & SSIM            \\ \hline
\multirow{2}{*}{NuFFT}     & 5 spokes / frame $(AF=25.6)$ & 21.58          & 0.3809          \\ 
                           & 13 spokes / frame $(AF=9.8)$ & 25.08          & 0.5250          \\ \hline
\multirow{2}{*}{GRASP}     & 5 spokes / frame $(AF=25.6)$ & 28.48          & 0.8429          \\ 
                           & 13 spokes / frame $(AF=9.8)$ & 29.11          & 0.8585          \\ \hline
\multirow{2}{*}{TD-DIP}    & 5 spokes / frame $(AF=25.6)$ & 29.24          & 0.8631          \\ 
                           & 13 spokes / frame $(AF=9.8)$ & 29.26          & 0.8735          \\ \hline
\multirow{2}{*}{HashINR}   & 5 spokes / frame $(AF=25.6)$ & 25.07          & 0.6257          \\ 
                           & 13 spokes / frame $(AF=9.8)$ & 27.73          & 0.7675          \\ \hline
\multirow{2}{*}{Ours}      & 5 spokes / frame $(AF=25.6)$ & 29.59 & 0.8712 \\ 
                           & 13 spokes / frame $(AF=9.8)$ & 30.13 & 0.8835 \\ \hline
\end{tabular}}
\end{table}

Tab.~\ref{tab:cardiac_5_13} reports the quantitative results analyzed on cardiac cine data. The reconstruction quality of NuFFT and HashINR is highly dependent on the number of spokes, resulting in substantial gaps in PSNR and SSIM values for $AF=25.6$ and $AF=9.8$, with differences ranging from 2.66 to 3.5 dB in PSNR and 0.1418 to 0.1441 in SSIM. 
In contrast, our results show relatively consistent reconstruction quality in both conditions by learning to reconstruct the canonical space in every iteration. Reconstructed images at specific time points are then obtained by warping the canonical space with the deformation field estimated based on temporal differences, leading to stable convergence in any condition. Our results achieve the best PSNR and SSIM values, 29.59 dB/0.8712 and 30.13 dB/0.8835 in $AF=25.6$ and $AF=9.8$, respectively.

\begin{figure*}[htb!]
\centering
\includegraphics[width=\textwidth, trim={0 1.4cm 0 0},clip]{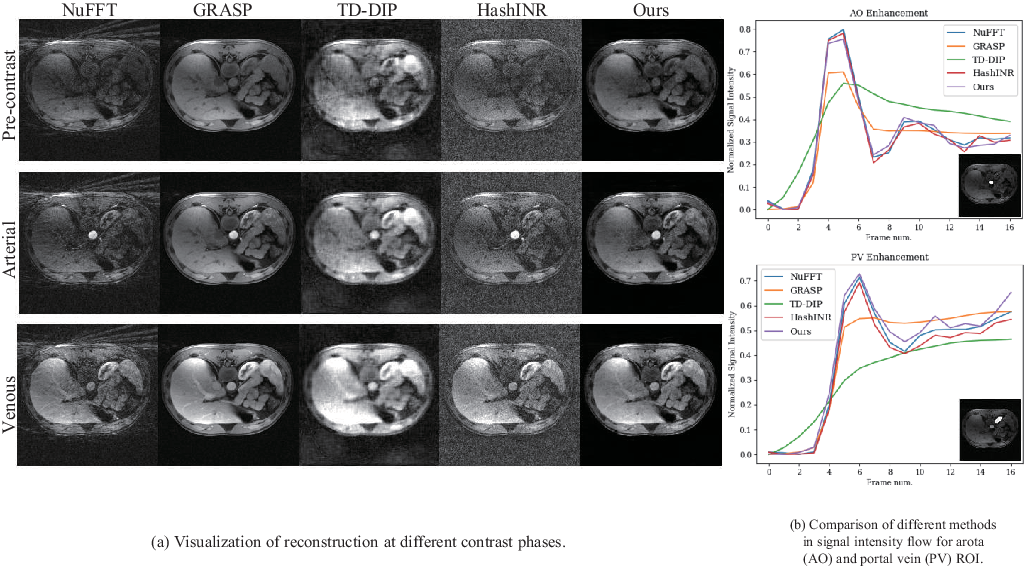}
\caption{Qualitative results of DCE liver data reconstruction with an undersampling ratio of 34 spokes per frame $(AF=11.3)$. We visualize reconstruction at different contrast phases (left), and compare signal intensity flow for aorta (AO) and portal vein (PV) ROI (right). 
}
\label{fig:liver_pro_recon}
\end{figure*}

\subsection{Dynamic Contrast-Enhanced (DCE) liver data}
Fig.~\ref{fig:liver_pro_recon} presents the qualitative reconstruction results and the corresponding signal intensity flowmap for DCE liver data reconstruction performed with an undersampling ratio of 34 spokes per frame $(AF = 11.3)$. GRASP reconstructs images in the $(y-x)$ domain with minimal noise, as shown (Fig.~\ref{fig:liver_pro_recon} (a)). However, it suffers from low temporal fidelity, as evident in the signal intensity flowmap (Fig.~\ref{fig:liver_pro_recon} (b)). On the other hand, HashINR exhibits good temporal fidelity in the flowmap, but its outputs display noticeable noise in the $(y-x)$ domain, overfitting to the undersampled frames. TD-DIP produces the reconstructions characterized by overly smooth appearances. This smoothness results in a failure to accurately delineate fine structural and contrast changes (Fig.~\ref{fig:liver_pro_recon} (a)). Consequently, it achieves the lowest temporal fidelity among the compared approaches, as reflected in the signal intensity flowmap in Fig.~\ref{fig:liver_pro_recon} (b). 
In contrast, our proposed method achieves a prominent performance, delivering both high temporal fidelity and phase-specific contrast changes. The reconstructions in Fig.~\ref{fig:liver_pro_recon} (a) are well-defined, showing clear structural details and accurate phase enhancements. Furthermore, our signal intensity flowmap in Fig.~\ref{fig:liver_pro_recon} (b) shows a strong capacity of our model to preserve temporal dynamics, capturing the changes in signal intensity over time with high accuracy. 

\subsection{Time consumption and GPU memory usage}
Table~\ref{tab:runtime_memory} presents the comparison of runtime and GPU memory usage of every method for dynamic MRI reconstruction at $AF=9.8$. 
GRASP requires 2.7 GB of GPU memory for cardiac cine data and 11.6 GB for DCE liver data because its cost on GPU memory depends on the image sequence size. TD-DIP utilizes the least GPU memory, but has the longest reconstruction time for cardiac cine data. HashINR takes 6.97 to 7.25 times longer runtime than ours, along with significantly higher memory consumption for both datasets. In contrast, our method achieves the shortest optimization time among learning-based methods for both data reconstruction with comparatively low GPU memory usage.

\begin{table}[t]
\caption{Runtime and GPU memory usage of different methods on each data, in the unit of seconds and gigabyte.}
\label{tab:runtime_memory}
\resizebox{\linewidth}{!}{%
\begin{tabular}{ccc|cc}
\hline
Data type & \multicolumn{2}{c|}{\begin{tabular}[c]{@{}c@{}}cardiac cine data\\ $128 \times 128$, 23 frames, 32 coils\end{tabular}} & \multicolumn{2}{c}{\begin{tabular}[c]{@{}c@{}}DCE liver data\\ $384 \times 384$, 17 frames, 12 coils\end{tabular}} \\ \hline
Method    & Runtime (sec)                & \begin{tabular}[c]{@{}c@{}}GPU memory\\ usage (GB)\end{tabular}                & Runtime (sec)              & \begin{tabular}[c]{@{}c@{}}GPU memory\\ usage (GB)\end{tabular}             \\ \hline
NuFFT     & 13.36                        & 0.7                                                                            & 18.96                      & 0.7                                                                         \\
GRASP     & 146.40                        & 2.7                                                                            & 423.24                      & 11.6                                                                        \\
TD-DIP    & 14164.31                      & 1.6                                                                            & 38581                      & 2.4                                                                         \\
HashINR   & 10484.32                     & 10.1                                                                           & 58088                      & 7.6                                                                         \\
Ours      & 1445.50(869.50)                       & 3.5                                                                            & 8329.55                      & 5.4                                                                           \\ \hline
\end{tabular}%
}
\end{table}

\begin{figure*}[htb!]
\centering
\includegraphics[width=\textwidth]{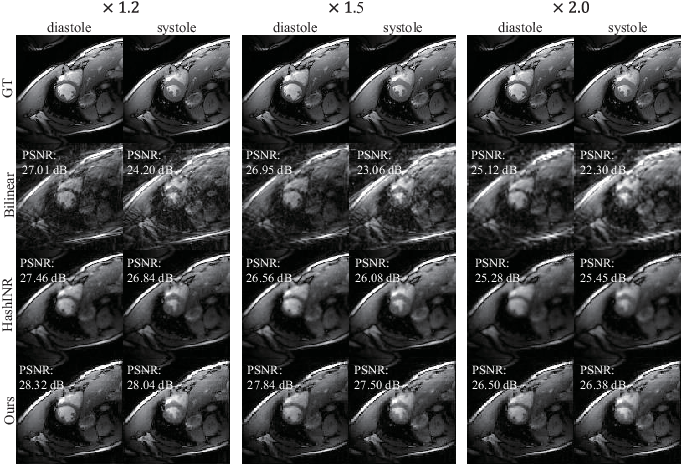}%
\caption{Qualitative results in spatial interpolation $(\times 1.2, \times 1.5, \times 2)$ task at $AF=9.8$. PSNR values are specific to each frame.}
\label{fig:spatialsr}
\end{figure*}

\begin{figure*}[htb!]
\centering
\includegraphics[width=\textwidth]{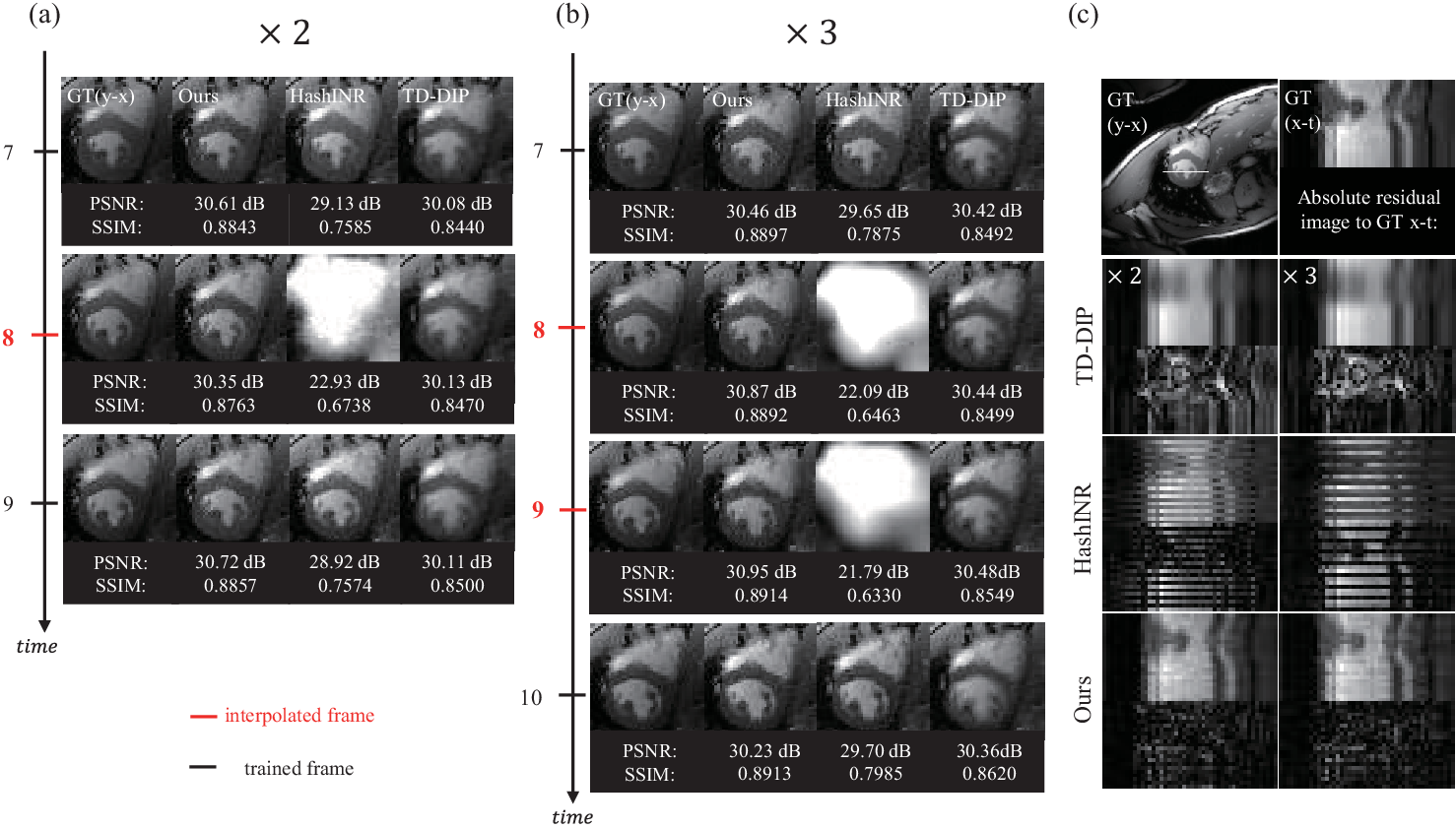}%
\caption{Comprehensive results in temporal interpolation $(\times 2, \times 3)$ task. (a), (b) Temporal interpolation results at $AF=6.1$ for $(\times2)$ and $(\times3)$, respectively. PSNR and SSIM values are specific to each frame. (c) The visual comparison of each method in the $(x-t)$ domain. A white line at the heart region of GT $(y-x)$ image indicates the cross section that is visualized.}
\label{fig:tempint}
\end{figure*}

\begin{table}
\centering
\caption{Quantitative results of spatial interpolation on cardiac cine data at $AF=9.8$.
}
\label{tab:sup_res}
\begin{tabular}{cccc}
\hline
            Method              & Scale ratio  & PSNR (dB) & SSIM   \\ \hline
\multirow{3}{*}{Bilinear} & $\times$1.2 (106 $\rightarrow$128) & 25.67     & 0.5758 \\
                          & $\times$1.5 (86 $\rightarrow$128) & 25.41     & 0.5785 \\
                          & $\times$2 (64 $\rightarrow$128)   & 24.28     & 0.5552 \\ \hline
\multirow{3}{*}{HashINR}  & $\times$1.2 (106 $\rightarrow$128) & 27.16     & 0.8053 \\
                          & $\times$1.5 (86 $\rightarrow$128) & 26.78     & 0.7960 \\
                          & $\times$2 (64 $\rightarrow$128) & 25.32     & 0.7464 \\ \hline
\multirow{3}{*}{Ours}     & $\times$1.2 (106 $\rightarrow$128) & 28.20     & 0.8516 \\
                          & $\times$1.5 (86 $\rightarrow$128)& 27.80     & 0.8323 \\
                          & $\times$2   (64 $\rightarrow$128) & 26.65     & 0.7929 \\ \hline
\end{tabular}%
\end{table}

\subsection{Results of spatial and temporal interpolation}
We conduct experiments of spatial interpolation $(\times 1.2, \times 1.5, \times 2)$ and temporal interpolation $(\times 2, \times 3)$ to validate the internal continuity of our model. For spatial interpolation, we employ bilinear interpolation, HashINR, and our proposed method. For temporal interpolation, we utilize HashINR, TD-DIP, and our method. 

For spatial interpolation, we optimize the models by taking as input spatial coordinates $(x,y)$ at time $t$, where $x,y \in [-1,1]$ is evenly undersampled according to the corresponding center-cropped \textit{k}-space acquisitions in the image sequence, and calculating loss between the given spokes derived from the center-cropped ones and the spokes from predicted ones. In inference, the spatial coordinates $(x,y)$ at time $t$, where $x,y \in [-1,1]$ is evenly spaced according to the original image size, are fed into the models for spatial interpolation of all frames. For evaluation, an undersampling ratio of 13 spokes per frame $(AF=9.8)$ is employed, with Peak Signal-to-Noise Ratio (PSNR) and Structural Similarity Index Measure (SSIM) used as the performance metrics. Fig.\ref{fig:spatialsr} shows qualitative results of spatially-interpolated reconstruction with an undersampling ratio of 13 spokes per frame $(AF=9.8)$. 
Applying bilinear interpolation directly to the undersampled data in the image domain results in significant noise and artifacts. The outputs generated by HashINR exhibit incremental blurriness, particularly around the papillary muscle, as the scale ratio increases. In contrast, our method effectively preserves structural details in all scale ratios. Furthermore, as shown in Tab.~\ref{tab:sup_res}, our approach consistently achieves the highest PSNR and SSIM values under all conditions. 

For temporal interpolation, we optimize the models by taking as inputs spatio-temporal coordinates corresponding to evenly subsampled frames according to the sampling ratio $(\times 2)$ or $(\times 3)$. During inference, the coordinates from all frames are fed into the models to generate whole image sequence. We extract the image features of two neighboring frames of time $t$ because we can not exploit the image features corresponding to the unseen time coordinates during optimization.
Fig.~\ref{fig:tempint} (a) and (b) show the overall results of reconstruction with an undersampling ratio of 21 spokes sampled per frame $(AF=6.1)$ and Fig.~\ref{fig:tempint} (c) visualizes output of each method in the $(x-t)$ dimension. 
As frame number goes from $7$ to $9$, the papillary muscle detaches from the wall in ground truth. The output of HashINR at time $9$ shows a clear detachment of the muscle from the wall, but the interpolated output at time $8$ completely loses structural integrity and its visual quality is extremely poor. As shown in Fig.~\ref{fig:tempint} (c), there are temporal vacancy artifacts in the outputs of HashINR in the $(x-t)$ domain. 
TD-DIP can not reconstruct the definite anatomy of the muscle and its contraction through the time. As a result, the outputs of TD-DIP are blurry and indistinct in the $(x-t)$ domain, as shown in Fig.~\ref{fig:tempint} (c).
However, our results show a distinct detachment of the muscle from the wall with the best quantitative and qualitative results (Fig.~\ref{fig:tempint} (a) and (b)). In addition, ours show the highest fidelity in the $(x-t)$ domain with less noise or artifacts with optimistic interpolation capacity (Fig.~\ref{fig:tempint} (c)).

\begin{table}
\centering
\caption{Quantitative results of ablation study on types of the encoder pairs. The ablation study is implemented on cardiac cine data at $AF=9.8$. $D$ denotes the deformation network and $C$ denotes the canonical network in our model.}
\label{tab:ablationstudy_enc_pairs}
\resizebox{\linewidth}{!}{%
\begin{tabular}{ccccc}
\hline
Encoder for $D / C$ & PSNR (dB) & SSIM   & \begin{tabular}[c]{@{}c@{}}GPU\\ memory\\ usage (GB)\end{tabular} & \begin{tabular}[c]{@{}c@{}}Runtime \\ (sec)\end{tabular} \\ \hline
Freq-Freq           & 29.00     & 0.8517 & 3.46                                                               & 1343.67                                                   \\ \hline
Hash-Freq           & 29.96     & 0.8895 & 3.48                                                               & 1587.43                                                   \\ \hline
Hash-Hash           & 29.84     & 0.8773 & 3.6                                                               & 1607.91                                                   \\ \hline
Freq-Hash (Ours)    & 30.13     & 0.8835 & 3.48                                                               & 1445.50                                                   \\ \hline
\end{tabular}%
}
\end{table}

\begin{table}
\centering
\caption{Quantitative results of ablation study on types of the image feature extractor. The ablation study is implemented on cardiac cine data at $AF=9.8$.}
\label{tab:ablationstudy_enc}
\resizebox{\linewidth}{!}{%
\begin{tabular}{ccccc}
\hline
        Feature Extractor                      & PSNR (dB) & SSIM   & \begin{tabular}[c]{@{}c@{}}GPU\\ memory\\ usage (GB)\end{tabular} & \begin{tabular}[c]{@{}c@{}}Runtime\\ (sec)\end{tabular} \\ \hline
w/o encoder                   & 29.59     & 0.8807 & 1.9                                                               & 1332.80                                                 \\ \hline
EDSR~\cite{edsr}              & 29.53     & 0.8790 & 3.5                                                               & 2826.45                                                 \\ \hline
RDN~\cite{rdn}                & 29.28     & 0.8750 & 12.0                                                              & 6024.91                                                 \\ \hline
SwinIR~\cite{liang2021swinir} & 29.34     & 0.8816 & 18.3                                                              & 5889.91                                                 \\ \hline
MDSR~\cite{mdsr} (Ours)       & 30.13     & 0.8835 & 3.5                                                               & 1445.50                                                  \\ \hline
\end{tabular}%
}
\end{table}

\subsection{Ablation study}
\subsubsection{Types of encoder pair}
We conduct an ablation study on different types of encoder pair: Freq-Freq, Hash-Freq, Hash-Hash, and Freq-Hash. Each encoder is attached to the deformation network and the canonical network, respectively. The effectiveness of these encoder pairs is evaluated on cardiac cine data with an undersampling ratio of 13 spokes per frame $(AF=9.8)$. All other conditions remain identical. 

In Tab.~\ref{tab:ablationstudy_enc_pairs}, the Freq-Freq pair, having no trainable parameters in its encoders, results in the least GPU memory usage. However, the reduction in memory usage compared to other encoder pairs is negligible, and it exhibits the poorest reconstruction quality. The Hash-Hash pair performs worse than the Hash-Freq pair or our proposed method, while requiring slightly more time to reconstruct a single cardiac cine dataset.

\subsubsection{Types of feature extractor}
We conduct an ablation study on different image feature extractors in our model: EDSR~\cite{edsr}, SwinIR~\cite{liang2021swinir}, RDN~\cite{rdn}, and MDSR~\cite{mdsr}. The effectiveness of these encoders is evaluated on cardiac cine data with an undersampling ratio of 13 spokes per frame $(AF=9.8)$. All other conditions remain identical. 
Tab.~\ref{tab:ablationstudy_enc} shows the quantitative results for each encoder type. While RDN and SwinIR consume over 10 GB of memory, their reconstruction quality is not even comparable to the result obtained without using an encoder. By contrast, our method achieves the best reconstruction with low GPU memory usage and requires only a short optimization time.

\section{Conclusion}
In this paper, we propose a novel framework for spatio-temporal representation learning tailored to dynamic MRI reconstruction without requiring ground truth data. Our method, Dynamic-aware HashINR (DA-INR), combines the efficiency of hash encoding for rapid optimization with an explicit design inspired by D-NeRF to effectively capture continuous temporal redundancy. By leveraging a canonical network, DA-INR incorporates temporal consistency into its structure, reducing dependency on explicit regularization terms while ensuring fast convergence. Comprehensive experiments demonstrate that DA-INR achieves superior reconstruction quality and efficiency, making it a robust solution for dynamic MRI reconstruction even under extreme undersampling conditions. 

\bibliographystyle{unsrt}  
\bibliography{sample}

\end{document}